\documentclass[iop]{emulateapj}

\usepackage{amsmath}
\usepackage{natbib}

\slugcomment{}

\newcommand{\be}{\begin{equation}}
\newcommand{\ee}{\end{equation}}

\shorttitle{SZ effect amplitude}
\shortauthors{L\'opez-Corredoira et al.}

\begin{document}

\title{Analysis of the Amplitude of the 
Sunyaev--Zel'dovich Effect out to Redshift $z=0.8$}

 \author{M. L\'opez-Corredoira \altaffilmark{1,2}, C. M. Guti\'errez \altaffilmark{1,2}, 
R. T. G\'enova-Santos \altaffilmark{1,2}}
\altaffiltext{1}{Instituto de Astrofisica de Canarias, E-38205 La Laguna, Tenerife, Spain; martinlc@iac.es}
\altaffiltext{2}{Departamento de Astrofisica, Universidad de La Laguna, E-38206 La Laguna, Tenerife, Spain}

\begin{abstract}
The interaction of the cosmic microwave background (CMB) with the hot gas in clusters of galaxies, the so-called Sunyaev--Zel'dovich (SZ) effect, is a very useful tool that allows us to determine the physical conditions in such clusters and fundamental parameters of the cosmological models. In this work, we determine the dependence of the the SZ surface brightness amplitude with redshift and mass of the clusters.
We have used PLANCK+IRAS data in the microwave-far infrared and a catalog with $\gtrsim 10^5$ clusters of galaxies extracted from the SDSS by Wen et al. (2012).
We estimate and subtract the dust emission from those clusters. From the residual flux, we extract its SZ flux densities.

The absolute value of the SZ amplitude indicates that the gas mass 
is around 10\% of the total mass for cluster masses of $M\sim 10^{14}$ M$_\odot $. 
This amplitude is compatible with no evolution with redshift and 
proportional to $M^{2.70\pm 0.37}$ (using X-ray derived masses) or 
$M^{2.51\pm 0.38}$ (using weak-lensing derived masses), with
some tension regarding the expectations of the self-similar dependence (amplitude proportional to $M^{5/3}$).

Other secondary products of our analysis include that clusters have a dust emission with emissivity index $\beta\sim 2$ and temperature $T\sim 25$ K; we confirm that the CMB temperature agrees with a dependence of $T_0(1+z)$ with clusters of much lower mass than those explored previously; and we find that 
the cluster masses derived by Wen et al. (2012) from a richness-mass relationship are biased by a factor of $(1+z)^{-1.8}$ with respect to the X-ray and weak--lensing measurements.

\end{abstract}

\keywords{galaxies: clusters: general --- cosmic background radiation}

\section{Introduction}

Galaxy clusters are the largest gravitationally bound structures in the universe and emerge in the cosmic web of large-scale structure at locations where the initial fluctuations created the largest potential wells (Allen et al. 2011). The baryon budget of these objects is dominated by a hot ionized intra-cluster medium (ICM), which is manifested in the X-ray range via thermal bremsstrahlung emission and also produces inverse Compton scattering of the cosmic microwave background (CMB) photons that is observed in the microwave range as a distortion of the CMB spectrum known as the Sunyaev--Zeldovich (SZ hereafter) effect (Sunyaev \& Zeldovich 1970, 1972). In the thermal SZ (tSZ) effect, the inverse Compton scattering is produced by the random motion of thermal electrons inside the ICM, whereas in the kinetic SZ (kSZ) effect it is produced by bulk motions of electrons. They can be distinguished thanks to the fact that the former has a very unique spectral dependency, while the latter presents a Planckian spectrum. The tSZ is proportional to the thermal electron pressure of the ICM along the line of sight, which is represented by the Compton parameter $y=(\sigma _{\rm T}/m_e c^2)\int n_{\rm e} T_{\rm e} dl$, where $n_{\rm e}$ and $T_{\rm e}$ are the electron number density and the electron temperature, $\sigma _{\rm T}$ is the Compton cross--section and $c$ is the speed of light. The main observable is proportional to the integrated Compton parameter over the cluster volume, $D_{\rm A}^2 Y_{\rm SZ} = (k_B\sigma_{\rm T}/m_{\rm e}c^2)\int n_{\rm e} T_{\rm e} dV$, where $D_{\rm A}$ is the angular--diameter distance to the cluster.

The SZ effect is a very useful tool not only from the point of view of galaxy-cluster physics but also of cosmology. The SZ signal integrated over the cluster extent, $Y_{\rm SZ}$, is related to the total thermal energy of the ICM gas, and therefore is proportional to the total cluster mass (Sunyaev \& Zeldovich 1972). On the other hand, the Compton parameter along the line of sight, $y$, is insensitive to surface brightness dimming, which makes the SZ effect a very efficient tool for identifying high-redshift clusters, which remain elusive in X-ray or optical surveys. In fact, in the last $\sim 15$ years, the field has evolved from the study of the tSZ effect in tens of previously known clusters (see, e.g., Carlstrom et al. 2002 and references therein), to the first discoveries of new objects in blind SZ surveys (Staniszewski et al. 2009) and, more recently, to the compilation of catalogs of many new SZ clusters, thanks to the improved angular resolution and/or wider frequency coverage of experiments like the Atacama Cosmology Telescope (Hasselfield et al. 2013), the South Pole Telescope (Reichardt et al. 2013), or Planck (Planck Collaboration 2016f). 

These properties make the SZ a very powerful cosmological tool. Not only does it allow for identification of high-redshift clusters but also, through the measurement of the integrated SZ fluxes, $Y_{\rm SZ}$, it allows for their masses to be estimated. By combining this information with redshift estimates, the cluster mass function (abundance of cluster per redshift interval and per mass interval) can be characterized. This function is very sensitive to the underlying cosmology, and therefore SZ observations can be used in this way to constrain cosmological parameters (Planck Collaboration 2016e). However, the use of the measured $Y_{\rm SZ}$ as a mass proxy requires a precise calibration of the $Y_{\rm SZ}-M$ relation, something that has been tackled both by using simulations (Nagain 2006, Bonaldi et al. 2007) and real observations (Bonamente et al. 2006, Melin et al. 2011). In an alternative way, cosmological parameters can also be constrained through the characterisation of the SZ effect power spectrum (e.g., Hasselfield et al. 2013; Reichardt et al. 2013; Planck Collaboration 2016d).

The combination of the tSZ effect with other observables, in particular, X-rays, have other applications from the point of view of cluster physics and cosmology. In fact, the different scaling of the SZ and X-ray fluxes with the electron temperature and with the electron density, make these two observables highly complementary. For this reason, the scaling between both has been extensively studied. Sometimes these studies have yielded scaling laws that are different from the expectations (Lieu et al. 2006, Komatsu et al. 2011), casting doubts on our correct understanding of cluster physics, whereas many other works (Atrio-Barandela et al. 2008; Melin et al. 2011; Planck Collaboration 2011; de Martino \& Atrio-Barandela 2016) have found results in nice agreement with the theory. The combination between the tSZ and X-rays also has a number of cosmological applications, like measuring the Hubble parameter  (Bonamente et al. 2006), probing the distance--duality relation between the angular-diameter and the luminosity distances (Uzan et al. 2004) or constraining the evolution of the CMB temperature with redshift (Luzzi et al. 2009, 2015; Hurier et al. 2014; de Martino et al. 2015).
Also, SZ measurements could be used to constrain the thermal history of the intra-cluster gas (Ferramacho \& Blanchard 2011).
Despite giving rise to a much weaker signal at the position of the richest galaxy clusters, the kSZ has the potential to be used to constrain bulk flows in the local universe (Kashlinsky \& Atrio-Barandela 2000). It has also already been used to measure pairwise momenta of clusters (Hand et al. 2012, Planck Collaboration 2014a), which has proven to be a powerful tool to explore the gas distribution around bright galaxies in the search for the missing baryons in the local universe (Hern\'andez-Monteagudo et al. 2015).

In this paper, we present a characterization of the SZ signal in the data of the Planck satellite (Planck Collaboration 2016a) toward the 132,684 galaxy clusters identified by Wen et al. (2012) in the Sloan Digital Sky Survey III (SDSS-III). The vast majority of these clusters are not sufficiently rich as to be detected individually in the Planck data. 
For this reason, we resort to a stacking technique in which we consider 
two binning schemes based on a cluster's mass (as estimated from optical richness) and redshift.
In Section 2, we describe the data used and, in Section 3, we explain the methodology for the determination of the SZ flux. In Section~4, we address the issue of removing the contamination induced by thermal cluster dust emission produced by galaxies in the
clusters distribution. In Section 5, we present the results of the fits of the SZ flux and of the CMB temperature with respect to the cluster mass and redshift and discuss the results. Finally, in Section 6, we summarize the main conclusions derived from this study.

\section{Data}

\subsection{Cluster Catalog}

We use the sample of clusters obtained by
Wen et al. (2012), from the SDSS (Sloan Digital Sky Survey)-III survey. 
The spatial coverage of that survey is $\sim 14,000$
square degrees and the catalog contains 132,684 clusters in the redshift range of
$0.05 \leq z \leq 0.80$. According to
those authors, the catalog is more than 95\% complete for clusters  with a mass of $M_{200} > 10^{14}$ M$_\odot$ in the range of $0.05 \leq z \leq 0.42$ and contains a false detection rate of less than 6\%. The catalog presents photometric redshifts for all the clusters, whilst only $\sim 30$\%  have determination of redshift based on the spectroscopy of their brighter cluster galaxy (an updated version
of the catalog with 52,683 spectroscopic redshifts was presented by Wen \& Han 2013). From these cases, it was estimated for the photometric redshifts, a systematic offset  $<0.004$ and a standard deviation $0.025-0.030$ for $z<0.45$ and $0.030-0.060$ for $z>0.45$.
Throughout the paper, we use the masses $M_{200}$ as estimated from optical richness, 
as the mass within the radius $r_{200}$ (the 
radius within which the mean density of a cluster is 200 times
the critical density of the universe), derived through (Wen et al. 2012):
\begin{equation}
\label{m200}
M_{\rm 200}=3.2\times 10^{12}\times {\rm Richness}^{1.17}\ {\rm M}_\odot
\end{equation}
The ``Richness'' is defined by $\frac{L_{200}}{L_*}$, where $L_*$
is the characteristic luminosity of galaxies in the r band, and
$L_{200}$ is the total r-band luminosity within $r_{200}$
in the SDSS $r$-band by summing luminosities of member galaxy candidates brighter than an evolution-corrected absolute magnitude $M^e_r(z)= M_r(z)+Q\,z\le -20.5$, with a passive evolution of $Q=1.62$. 

In Eq. (\ref{m200}), the exponent has an error bar of $\Delta e=0.03$ (Wen et al. 2012), which means that the correct masses $M_{200}^{\rm corr.}$ are related to the calculated $M_{200}$ by: $M_{200}^{\rm corr.}=M_{200}\times 
\left(\frac{M_{200}}{3.2\times 10^{12}\ {\rm M_\odot}}\right)^{\Delta e/e}$.
The error in the normalization of the mass might be important, but the dependence
with the mass has only an exponent of $\lesssim 0.025$, which is unimportant
for the present calculations. However, Ford et al. (2014) give an exponent 
$1.4\pm 0.1$ instead of $1.17\pm 0.03$. If there
was an error of $\Delta e\approx 0.2$, we would have an extra dependence proportional to $M_{200}^{0.17}$, which is more important, but it will not wreck the main results obtained in this paper: adopting a Ford et al.
(2014) exponent would imply that we would need to substitute along this paper
$M_{200}$ for $M_{200}^{0.85}$.

To extend the studies to lower redshifts, we also use the sample of 1059 Abell clusters with available spectroscopic redshifts.\footnote{\tt http://heasarc.gsfc.nasa.gov/W3Browse/galaxy\-catalog/abellzcat.html}

Our analysis uses a cluster catalog that covers masses much lower than those of the PLANCK catalog. Wen et al. clusters have also been used by the Planck Collaboration (2016d), but in a more restrictive way, without paying attention to the CMB temperature or the dependence on the redshift and without the more careful dust substraction that we have carried out here.

\subsection{Planck+IRAS data}

The Planck satellite (Planck Collaboration 2016a) was an ESA mission launched in 2009 that surveyed the full sky in nine independent frequency bands provided by the LFI (30, 44, and 70~GHz) and by the HFI (100, 143, 217, 353, 545, and 857~GHz) instruments, and with better angular resolutions as compared to previous surveys in this range of frequencies between 33~arcmin (at 30 GHz) and 5~arcmin (at 217--857~GHz). These characteristics make it an ideal observatory not only to study the CMB anisotropies, but also the SZ effect in galaxy clusters and many other topics. Its frequency coverage allows us to trace the negative ($\nu< 217$~GHz) and the positive ($\nu> 217$~GHz) parts, as well as the null ($\nu= 217$~GHz) of the SZ spectrum. Furthermore, its fine angular resolution allows us to explore high-redshift clusters. This implied a great improvement over WMAP
(Bennett et al. 2013), the previous NASA mission which, with a highest frequency of 94~GHz and finest angular resolution of 13.2~arcmin only allowed us to study the SZ effect in nearby and rich clusters. Another important asset of Planck wide frequency coverage is that it permits a much better removal of foreground contaminants.

We have used the DR2 data release (Planck Collaboration 2016a) in the form of full-sky maps projected in Healpix pixelization (G\'orski et al. 2005), downloaded from the Planck Legacy Archive,\footnote{\tt http://www.cosmos.esa.int/web/planck/pla} and with pixel resolutions of $N_{\rm nside}=1024$ (pixel size of 3.4~arcmin) for the LFI and $N_{\rm nside}=2048$ (pixel size of 1.7~arcmin) for the HFI. In the LFI, we have used only the 70~GHz frequency band. The lower frequencies were dropped from the analysis owing to their coarse angular resolution. On the other hand, we use the six HFI frequency bands, the highest ones being very important for subtraction of the thermal dust component. We select the data within the region outside of mask 2 of PLANCK, 40\% of the sky, which avoids the low Galactic latitude regions and other lines of sight with high Galactic dust columns (Planck Collaboration 2014b). By removing the lowest latitude regions, apart from excluding the highest Galactic dust regions, we also exclude the fields with higher foreground of CO.
The foreground and primary CMB anisotropies are not correlated with the positions of clusters, they only introduce uncorrelated noise in the analysis (we carry out some experiments using the Planck PR1 CMB map to check it); CMB anisotropies would be important if we analyzed individual clusters, but for a stacking of clusters this noise is quite low, so we do not remove them.

To extend the spectral range analyzed
and to better constrain the spectral shape of the dust function
emission, we  also used the IRAS map at 100 $\mu $m. In particular, we used
the new generation of IRAS processed images, which incorporates
improvement on the zodiacal light subtraction, on destripping and have a
zero level compatible with COBE-DIRBE. The generation of the map is described by Miville-Deschenes \& Lagache (2005; IRIS maps). For this map, we also
used the maps in HEALPIX (G\'orski et al. 2005) format with resolution
$N_{\rm nside}=2048$.

\section{Flux measurements}

\subsection{Stacked samples}

Due to the low emission of a single cluster, we used a statistical
approach in which the emission of all the clusters within given ranges of redshift 
and masses (as estimated from optical richness by Wen et al. 2012) 
are stacked around their centers and 
averaged (see Guti\'errez \& L\'opez-Corredoira 2014 for a
full description of the method).

Possible asymmetries or irregularities
in each cluster are randomly oriented and independent between clusters,
so stacking the emission from many clusters will largely smooth those  asymmetries,
and then in the stacked map we can ignore any angular dependence and
consider only the radial profile. Two different  binning methods (denoted
through this paper as modes A and B) were used. In A, the ranges in redshift
and masses were  selected in order to have roughly the same number of
clusters in each bin. In B, we defined bins with the same width in
redshift (0.12) and mass (0.3 dex). The number of low-mass or low-redshift clusters  is comparatively higher than those of high-mass or
high-redshift clusters (see Tables \ref{Tab:binsa}, \ref{Tab:binsb}). 
In the catalog by Wen et al. (2012), this procedure
produces bins with a very different number of objects.

\subsection{Sky subtraction and flux integration}

A sky subtraction is done by subtracting for each stacked map of clusters the average
background. Galactic dust was not previously subtracted. The subtraction of a Galactic dust foreground would also partially or totally subtract the dust emission of the cluster (without K-correction) and we prefer to subtract separately later, taking into account the 
redshift of the clusters.

\begin{figure}
\vspace{.5cm}
\centering
\includegraphics[width=8cm]{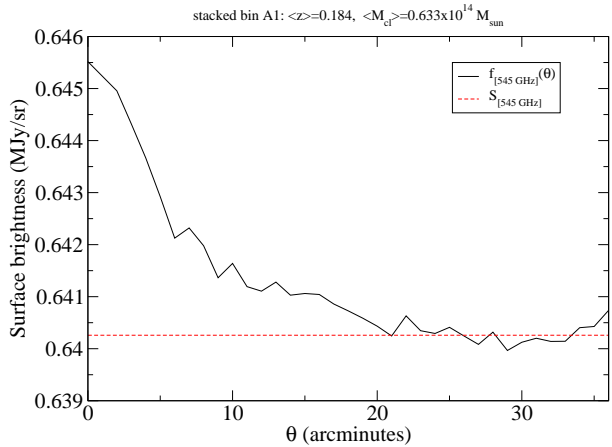}
\caption{Average measured surface brightness per cluster as a function of the distance to the center of the cluster for bin A1 at 545 GHz and its corresponding sky emission to be
subtracted.}
\label{Fig:bin1_skysubs}
\end{figure}

We proceed to apply an aperture photometry of the stacked maps for each frequency $\nu $
(see Fig. \ref{Fig:bin1_skysubs}). For all clusters, we measure the flux in the central 
$\theta _{\rm max}=10.5'$, which gathers most of the flux of the clusters. The procedure is as follows.

First, for each of the bins, we calculate the surface brightness $f_\nu(\theta )$ in the stacked maps with rings between angular distances $\theta $ and $\theta +d\theta $ (we use $d\theta =1'$), where $R$ is the angular distance to the cluster center. Then, we calculate the average sky emission ($S_{{\rm sky}, \nu }$) as the weighted average of surface brightness in rings between $\theta =20'$ and $\theta =36'$, where the weight is proportional to the number of pixels in each ring. Cluster emissions are within $\theta <20'$, and a sky ring close to the cluster is preferable to minimize the effect of the average gradients of sky emission; that is the reason for the radii selection. With that, we get a first profile of the cluster: $f_{1,\nu}(\theta )=f_\nu (\theta )-S_{{\rm sky}, \nu }$. An example is given in 
Fig. \ref{Fig:bin1_skysubs} for the stacked bin A1 in the frequency 545 GHz.

Several clusters may be observed in a given line of sight. 
This is corrected by the method explained in appendix \ref{.sevclu}.
Also, the loss of flux due to beam dilution is corrected with the method explained in 
appendix \ref{.corrfluxloss}

The fluxes $F_\nu$ obtained by this method were compared for some single clusters detected
also by Planck to the values given by Luzzi et al. (2015), and we get very similar results.
For instance, for the cluster Abell 85, we get the fluxes of -7, -6, -4, 0 , +10 $\mu $K arcmin$^2$
respectively at 70, 100, 143, 217, 353 GHz, similar to those obtained by 
Luzzi et al. (2015, Fig. 4 right) within the error bars (typically less than 20-30\% for each point, except for the point of 70 GHz in which the error is larger).

\section{Estimation and subtraction of the thermal dust component}

The Galactic dust produces an even higher contamination, but this is uncorrelated with the position of the clusters, so this only produces noise in the stacked maps. 
Also the extragalactic sources uncorrelated with the cluster produce noise
for the same reasons.
However, clusters have important dust emission (Guti\'errez \& L\'opez-Corredoira 2014, 2017; Planck Collaboration 2016g) and this should be
subtracted in order to examine the Sunyaev--Zel'dovich effect.
A fit to the cluster dust is better than including this emission as part of the Galactic dust because this takes into account the K-correction due to the redshift of the emitted flux. 

To subtract the signal of this dust, we follow this procedure: with the flux $F_\nu$ for $\nu $=545, 857, and 2998 GHz for a given stacked image of clusters with
average redshift $\langle z\rangle$, we fit the three parameters ($A_{\rm dust}$, $\beta _{\rm dust}$, $T_{\rm dust}$) of a dust emission function given by
\begin{equation}
\label{fdust}
F_\nu =F_{\rm dust}(\nu )=A_{\rm dust}\,\nu_{rest}^{\beta _{\rm dust}}\,B(\nu_{rest},T_{\rm dust})
,\end{equation}\[
\nu_{rest}=\nu (1+z)
,\] 
where $B(\nu ,T)$ is the black body emission function, $\beta _{\rm dust}$ is the emissivity, and $A_{\rm dust}$ is the amplitude.
We have chosen the frequencies 545, 857, and 2998 GHz (100 $\mu $m) because the contribution of the Sunyaev--Zel'dovich
effect here is negligible and a possible second component of dust with higher temperature would have only 
a negligible contribution. We do not use the 4997 GHz (60 $\mu $m) of IRAS because this would need a second
temperature of dust to be fitted (Xilouris et al. 2012).
Although there is some degeneracy in the space $\beta _{\rm dust}$-$T_{\rm dust}$ in the search of the best fit, allowing the variation of both parameters improved the fit. This is the reason for that choice.
The error bars of the parameters are derived with the standard $\chi ^2$ analysis (Avni 1976).

At first, we neglect the contribution of Sunyaev--Zel'dovich at 545 GHz for a first fit of the dust. Later, once we have the Sunyaev--Zel'dovich fit according to \S \ref{.fitSZ}, we subtract the contribution of Sunyaev--Zel'dovich at 545 GHz and we redo the fit of dust. At 857 and 2998 GHz, we always consider the Sunyaev--Zel'dovich effect to be negligible. One may wonder why we do not fit the dust and Sunyaev--Zel'dovich together at the same time: the answer is that, in such a case we would be dominated by high-frequency points, which get higher signal/noise, and the Sunyaev--Zel'dovich effect would get a poorer fit. Therefore, we do the fit at high frequencies for dust and low frequencies for 
Sunyaev--Zel'dovich separately.

Tables \ref{Tab:binsa} and \ref{Tab:binsb} give values for the bets fits of $\beta _{\rm dust}$ and $T_{\rm dust}$, which are within the normal values of the fits in other galaxies (Shetty et al. 2009; Galametz et al. 2012).
We have tried the dust subtraction by stacking clusters lying in different stripes of constant Galactic latitude, using masks 1 or 2 of PLANCK, covering, respectively the 20\% and 40\% lowest extinction regions of the sky respectively. The results do not have significant differences. We choose here the values of mask 2 because they provide lower error bars and they cover a region of the sky that is still very conservative in avoiding the high extinction regions. 
This is also in agreement with the analysis of the dust in clusters
with PLANCK by Planck Collaboration (2016g).

\subsection{Effective frequencies}
\label{.efffreq}

A minor correction is carried out by recalculating the effective frequencies of each of the filters, rather than taking the nominal value.

For each of the filters, we calculate the frequency averaged on the detected flux convolved with the filter transmission as:
\begin{equation}
\nu _{\rm eff.}(\nu )=\frac{\int _0^\infty d\nu '\,\nu '\,T_\nu(\nu ')\,F_{\rm dust}(\nu ')}
{\int _0^\infty d\nu'\,T_\nu(\nu ')\,F_{\rm dust}(\nu ')}
,\end{equation}
where $T_\nu(\nu ')$ is the transmission function of the filter with nominal frequency $\nu $.
Once we have these new frequencies $\nu _{\rm eff.}(\nu )$, we set the values of the 
fluxes $F_{\nu _{\rm eff.}(\nu )}$ as equal to the previously measured $F_\nu $ and we redo the
fit given by Eq. (\ref{fdust}), obtaining new parameters $A_{\rm dust}$, $\beta _{\rm dust}$,
$T_{\rm dust}$ that take into account this correction of frequencies. Those are the values
that we use.

\begin{figure}
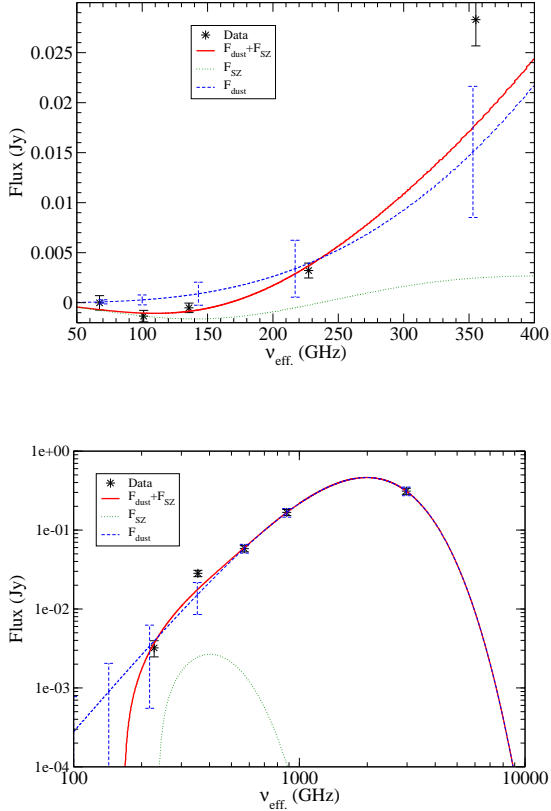

\vspace{.5cm}
\centering
\includegraphics[width=7.2cm]{bin1_fit.eps}
\\ \vspace{1cm} 
\includegraphics[width=7.2cm]{bin1_fitb.eps}
\caption{Linear and log--log plots of the flux vs. frequency with the corresponding fits of dust and
Sunyaev--Zel'dovich contribution for the stacked bin A1. The error bars of the dust fit correspond
to the error to its extrapolation (see the text).}
\label{Fig:bin1fit}
\end{figure}

\section{Fit of thermal Sunyaev--Zel'dovich effect}
\label{.fitSZ}

Once we estimate the dust contribution, we get the residuals 
$F_{\rm SZ}(\nu )=F_\nu-F_{\rm dust}(\nu)$. 
Kinematic Sunyaev--Zel'dovich (kSZ) is considered to be negligible here. 
Given that the kSZ signal has an opposite sign for clusters with positive or negative radial peculiar velocity, the stack of many clusters in a given redshift or mass bin will lead to an effective cancellation of this signal.
This extrapolation of the dust flux to lower frequencies involves an error given
by the $\chi ^2$ analysis of the dust fit, i.e. Error$[F_{\rm dust}(\nu)]$ is such that
there is a set of parameters necessary to produce a deviation Error$[F_{\rm dust}(\nu)]$
for the flux in the extrapolation of $F_{\rm dust}$ at frequencies $\nu $ which
gives an excess of $\chi ^2$ equal to 3.53 (corresponding to a fit with three free parameters;
Avni 1976). This error of the extrapolation is very important, becoming even larger than
the error of the flux measurement, as it can be observed in the example of Fig. \ref{Fig:bin1fit}.

The prediction at 353 GHz is obtained by extrapolating  the fits from the higher frequencies and may be underestimated  due to the possible  existence of a colder component of dust.
This could explain the poor fit at this frequency.

We neglect the possible emission of the clusters that could come from
CO lines of 115 GHz, 230 GHz, and 345 GHz (Planck Collaboration 2014b), because we are binning
sources with different redshifts, which will not sum the lines in a given redshift but distribute the CO light along a wide range frequencies, making its detection negligible.
Free-free emission of the cluster is more important at the lowest frequencies, but 
negligible within the error bars of our measurements (Luzzi et al. 2015).

Then, we fit the thermal Sunyaev--Zel'dovich emission/absorption to these residuals 
(Planck Collaboration 2016d; Sembolini et al. 2013; Luzzi et al. 2015):

\begin{equation}
\label{fsz}
F_{\rm SZ}(\nu )=A_{\rm SZ}\,\frac{T_{\rm CMB}(z)}{(1+z)}\frac{\nu ^2\,x^2\,e^x}{(e^x-1)^2}\,\left[x\,\coth \left(\frac{x}{2}\right)-4\right]
,\end{equation}\[
x=\frac{h\,\nu (1+z)}{k_B\,T_{\rm CMB}(z)}
,\]\[
A_{\rm SZ}=(0.03072385\,{\rm MJy}\,{\rm sr}^{-1}\,{\rm K}^{-1}\,{\rm GHz}^{-2})
\frac{k_B\,\sigma _T}{m_e\,c^2}
\]\[
\times \int_0^{\theta_{\rm max}}d\theta\,\theta\int _0^\infty ds\,n_e\,T_e
\]\[
=(0.03072385\,{\rm MJy}\,{\rm K}^{-1}\,{\rm GHz}^{-2})
\frac{k_B\,\sigma _T}{m_e\,c^2\,D_A(z)^2}
\]\begin{equation}\times
\int_{{\rm cluster}}dV\,n_e\,T_e
\label{asz}
,\end{equation}
where $h$ is the Planck constant, $k_B$ is the Boltzmann constant,
$m_{\rm e}$ is the electron mass, $\sigma _T$ is the Thomson cross-section, $n_{\rm e}$ is
the electron number density, $T_{\rm e}$ is the electron temperature, $s$ the distance along
the line of sight, $V$ is the physical volume in the cluster region, and $D_A$ is its angular distance. Note that $A_{\rm SZ}$ and $T_{\rm CMB}$ are the two free parameters
of the fit.

Note that in Eq. (\ref{fsz}) the spectral dependency of the SZ effect does not include the relativistic corrections (Itoh et al. 2002). Accounting for this correction would require knowledge of the electron temperatures of all galaxy clusters. This is the reason why this effect is not considered in other similar works (Bonamente et al. 2008, Melin et al. 2011, Planck Collaboration 2011, Hurier et al. 2014, de Martino \& Atrio-Barandela 2016). However, the correction is expected to be very small. Luzzi et al. (2015), who use a smaller sample with $T_{\rm e}$ measurements and therefore can characterize this effect, conclude that the relativistic correction affects their $T_{\rm CMB}$ estimated at the 2\% level for very rich clusters ($T_{\rm e}> 11$~keV). In our case, because we have much
fewer massive clusters on average, the correction will be much lower than 2\%. Also, Hurier et al. (2014) conclude that the relativistic corrections should have a very minor effect on their analyses, and they argue that they can safely neglect them. This also applies to our sample because Hurier et 
al. (2014) do not see any correlation of $T_{\rm e}$ with $T_{\rm CMB}$, 
and the range of redshifts and masses of their and our cluster samples are similar.

The flux $A_{\rm SZ}$ may be related to a luminosity $L_{\rm SZ}$ as
\begin{equation}
\label{lsz}
L_{\rm SZ}=4\pi D_L(z)^2 A_{\rm SZ}
\end{equation}\[
\ \ \ \ =(4.331\times 10^{-59}\,{\rm W}\,{\rm K}^{-2}\,{\rm GHz}^{-2})
N_e\,\langle T_e\rangle (1+z)^4
,\]
where $N_e$ is the total number of electrons in the cluster, and
$D_L=(1+z)^2D_A$ is the luminosity distance.
For the calculation of the angular distance, we use a standard cosmology with
$h=0.70$, $\Omega _m=0.30$, $\Omega _\Lambda =0.70$.

In the cases where we get negative $A_{\rm SZ}$, we do not fit any $T_{\rm CMB}$; note, however, that they are used for the fit of the amplitude, so we do not have any bias in this. Like in \S \ref{.efffreq}, we calculate the effective frequencies and we redo the fit with these new frequencies.
That is, we carry out first the calculation of $F_{\rm SZ}(\nu)$ 
with the effective frequencies derived from PLANCK bandpasses (o PLANCK spectral responses) assuming a flag spectrum and, later, we calculate the effective frequencies $\nu_{\rm eff}$ using the fitted model.
The error bars of the parameters are derived with the standard $\chi ^2$ analysis (Avni 1976).
The results of the fits are given in Tables \ref{Tab:binsa} and \ref{Tab:binsb}.

\subsection{CMB temperature}

\begin{figure}
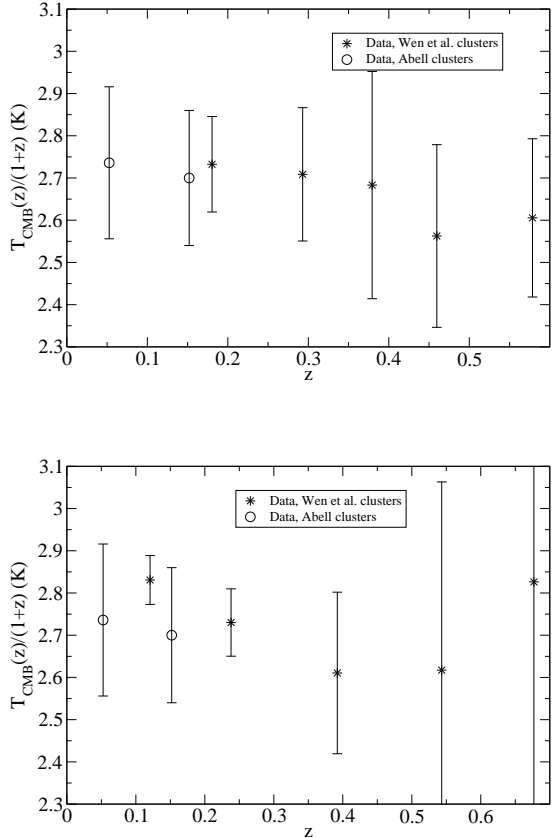

\vspace{.5cm}
\centering
\includegraphics[width=7.2cm]{SZdep4.eps}
\\ \vspace{1cm} 
\includegraphics[width=7.2cm]{SZdep4b.eps}
\caption{Dependence on redshift of $T_{\rm CMB}$ for bins of mode A (top) and mode B (bottom).}
\label{Fig:tempfits}
\end{figure}

The analysis of the CMB temperature is not the main aim of this paper. Nonetheless, with
the present numbers of tables \ref{Tab:binsa} and \ref{Tab:binsb}, we could see that
our data are compatible with the standard model prediction $T_{CMB}=2.726(1+z)$ K. 
In particular, we fitted our data (including Abell clusters) to 
$T_{CMB}=T_0(1+z)^\beta$, with $T_0=2.78\pm 0.14$, $\beta =-0.14\pm 0.19$ for Mode A, and 
$T_0=2.88\pm 0.11$, $\beta =-0.25\pm 0.22$ for Mode B (see Fig. \ref{Fig:tempfits}).
This confirms previous results obtained from the Sunyaev--Zel'dovich effect on CMBR (Luzzi et al. 2009, 2015; Avgoustidis et al. 2012; Saro et al. 2014; Hurier et al. 2014; de Martino et al. 2015) or with absorptions in the line of sight of quasars (Molaro et al. 2002, Noterdaeme et al. 2011, Muller et al. 2013). Therefore, our analysis is a corroboration of previous results using clusters with much lower masses than in previous papers, though
our errors are worse since we have not optimized our data selection for the
analysis of $T_{\rm CMB}$.

\subsection{Dependence of the amplitude on the mass and the redshift}

We bin the data with the same redshift and different masses (as estimated from optical richness), or with the (approximately) same mass and different redshifts, doing weighted averages in all cases (the weight of each
point is inversely proportional to the square of the error bar).
We use different functional shapes to explore the best fits:
\begin{equation}
\frac{L_{\rm SZ}}{(1+z)^4}=L_1\,\left(\frac{M_{\rm 200}}{10^{14}\,M_\odot}\right)^{\alpha _1}
,\end{equation}\begin{equation}
\frac{L_{\rm SZ}}{(1+z)^4}=L_2\,(1+z)^{\alpha _2}
.\end{equation}

\begin{figure}
\vspace{.5cm}
\centering
\includegraphics[width=7.2cm]{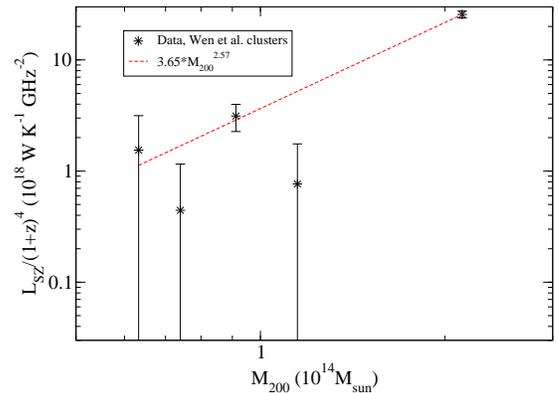}
\\ \vspace{1cm} %
\includegraphics[width=7.2cm]{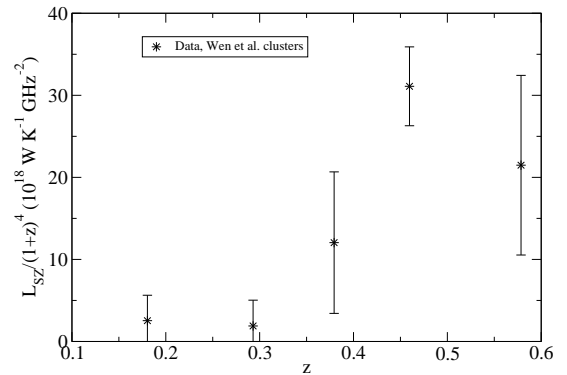}
\caption{Dependence on mass (as estimated from optical richness by Wen et al. 2012) and redshift of $L_{\rm SZ}$ for bins of mode A.
Best fits are plotted when they exclude a constant dependence within 2$\sigma $.}
\label{Fig:SZdep}
\end{figure}

\begin{figure}
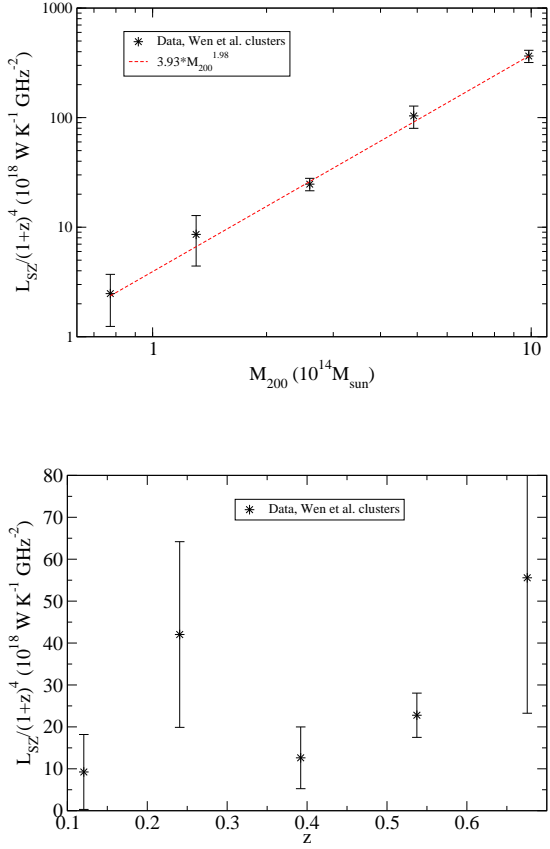

\vspace{.5cm}
\centering
\includegraphics[width=7.2cm]{SZdep1b.eps}
\\ \vspace{1cm} %
\includegraphics[width=7.2cm]{SZdep3b.eps}
\caption{As in Figure \ref{Fig:SZdep}, but for the bins of mode B.}
\label{Fig:SZdepb}
\end{figure}

The dependence on mass and redshift of both variables is shown in Figs. \ref{Fig:SZdep} and
\ref{Fig:SZdepb}.
We include in the calculation Error(z)=$\frac{r.m.s.(z)}{\sqrt{N_{cl}}}$.
Also, we have tried a fit with the 25 points of Wen et al. clusters without binning, with weights,
and a double dependence of the shape: 
\begin{equation}
\label{doublefit}
\frac{L_{\rm SZ}}{(1+z)^4}=L_3\,\left(\frac{M_{\rm 200}}{10^{14}\,M_\odot}\right)^{\alpha _3}(1+z)^{\alpha _4}
,\end{equation}
The obtained numbers of these fits are in Table \ref{Tab:fits}.

\begin{table}
\caption{Parameters of the dependence on mass $M_{200}$ (as estimated from optical richness by Wen et al. 2012)
and redshift of $L_{\rm SZ}$ (units: $10^{18}$ W K$^{-1}$ GHz$^{-2}$)
for bins of mode A and mode B.}
\begin{center}
\begin{tabular}{ccc}
Parameter & Mode A & Mode B  \\ \hline
$L_1$ & $3.65\pm 0.81$ & $3.93\pm 0.77$ \\
$\alpha _1$ & $2.57\pm 0.30$ & $1.98\pm 0.12$ \\
$L_2$  & $1.78\pm 2.64$ & $12.0\pm 10.2$ \\ 
$\alpha _2$ & $7.24\pm 3.91$ & $1.83\pm 2.12$ \\
$L_3$ & $2.81\pm 1.13$ & $5.14\pm 2.41$\\ 
$\alpha _3$ & $1.66\pm 0.35$ & $1.78\pm 0.22$ \\
$\alpha _4$ & $3.90\pm 0.79$ & $1.49\pm 1.35$ \\ \hline
\end{tabular}
\end{center}
\label{Tab:fits}
\end{table}

\begin{figure}
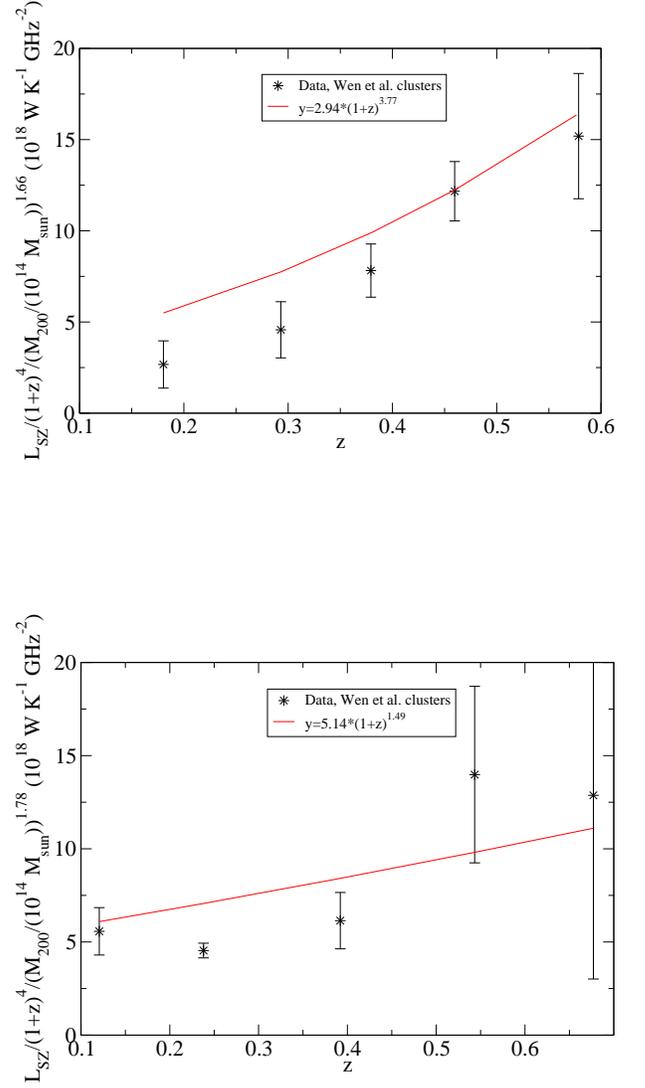

\vspace{1cm}
\centering
\includegraphics[width=8cm]{lsz5.eps}
\\
\vspace{2cm}
\includegraphics[width=8cm]{lsz5b.eps}
\caption{Dependence on redshift of $\frac{L_{\rm SZ}}
{(1+z)^4\left(\frac{M_{\rm 200}}{10^{14}\,M_\odot}\right)^{\alpha _3}}$ for bins of modes A and B respectively. The solid lines do not correspond to fits of these points, but
to the fits of $\alpha _4$ of Table \ref{Tab:fits}.}
\label{Fig:lsz5}
\end{figure}

There are two dependences with detection $>2\sigma $:

\begin{enumerate}

\item There is a dependence of the amplitude luminosity with $M_{200}$. We obtain power 
laws with exponents $\alpha _1$ or $\alpha _3$ between 1.66 and 2.57, being the best measurement with
lower error $1.98\pm 0.12$ ($\alpha _1$ for Mode B), which is not far (though still at 2--3$\sigma $) from
the self-similar value (5/3; Kaiser et al. 1986), from what is expected for this mass range according to simulations (Bonaldi et al. 2007; Sembolini et al. 2013, 2014), 
and from what is obtained with observations (e.g. Andersson et al. 2011; Sereno et al. 2015; 
Saliwanchik et al. 2015).
This value of 5/3 stems from $L_{SZ}\propto N_e\langle T_e\rangle$ [Eq. (\ref{lsz})],
and $N_e\propto M_{\rm gas}\propto M_{200}$, $\langle T_e\rangle \propto M_{200}^{2/3}$ (Sembolini et al. 2013). With the same data and different analysis by the Planck Collaboration (2016d, \S 4.5) an exponent of $1.92\pm 0.42$ is obtained, which is a similar value to ours, though our error bars are much smaller.
Indeed, the Planck Collaboration (2016d, \S 4.5) used a slightly different mass proxy based on the same data: $M_{200}^{\rm Wen-Planck}=(1.1\times 10^{14}\ {\rm M}_\odot)\left(\frac{N_{200}}{20}\right)^{1.2}$,
where $N_{200}$ is the number of galaxies brighter than a given magnitude within radius $R_{200}$, instead of our Equation (\ref{m200}). With a fit for all of the clusters of Wen et al. (2012), we can see that both amounts are related, on average, as $M_{200}^{\rm Wen-Planck}=0.77\,M_{200}^{0.918}(1+z)^{-0.833}$; if we apply this relationship, we get for Mode B $\alpha _1=1.94\pm 0.20$ (instead $1.98\pm 0.12$).

\item There is a dependence of $\frac{L_{\rm SZ}}{(1+z)^4}$ with redshift, once we discount the dependence on mass ($\alpha _4$) for mode A; in mode B, there is also some signal but it is less significant.
This is also reflected in the exponent $\alpha _2$, but here it is not significant and
part of the dependence might be due to a different relative weight of high-mass cluster at higher redshifts.
We plot $\frac{L_{\rm SZ}}{(1+z)^4\left(\frac{M_{\rm 200}}{10^{14}\,M_\odot}\right)^{\alpha _3}}$ versus redshift, assuming the measured value of $\alpha _3$, in Fig. \ref{Fig:lsz5}. There is clearly a trend to increase $\frac{L_{\rm SZ}}{(1+z)^4}$ for a given cluster mass with redshift, at the $5\sigma$ level for mode A. 
The differences of the points with the fit are most likely to irregularities in the mass dependence at low mass.
 
\end{enumerate}

The dependence of $L_{SZ}/(1+z)^4$ on mass is expected, but its
dependence with the redshift is not. According to Eq. (\ref{lsz}), we see that 
$L_{\rm SZ}$ is proportional to $N_e\langle T_e\rangle$. Theoretical analyses of these dependences, assuming hydrostatic equilibrium and isothermal distribution for dark matter and gas particles
within standard cosmology,
give $N_e\propto M_{\rm gas}$ and $T_e\propto [M_{\rm 200}E(z)]^{2/3}$, where
$E(z)=\sqrt{\Omega _m(1+z)^3+\Omega _\Lambda }$; hence $\frac{L_{\rm SZ}}{(1+z)^4}
\propto f_{\rm gas} M_{200}^{5/3}E(z)^{2/3}$ (Sembolini et al. 2013). 
The factor $E(z)^{2/3}$ would be between 1.0 and 1.3 for redshifts
between 0 and 0.7, which is much lower than the average factor $(1+z)^{\alpha _4}$ with
$\alpha _4\approx 4$, which is between 1.0 and 8.3 for redshifts between 0 and 0.7.
The simulations using Marenostrum-MultiDark SImulations of galaxy Clusters (MUSIC; Sembolini et al. 2013) do not predict either any of our observed evolution.
Nonetheless, Sembolini et al. (2013) claim to observe in their simulations a more conspicuous
evolution for lower masses, which is in line with what we observe; the difference is that
they see very little evolution in $\frac{L_{\rm SZ}E(z)^{-2/3}}{(1+z)^4}$ and we see a factor of $\sim 6$ for that quantity between $z=0$ and $z=0.7$.

The photometric redshifts have some small errors for each individual cluster 
(statistical errors of $\sigma _z=0.025-0.030$ for $z<0.45$ and $\sigma _z=0.030-0.060$ for $z>0.45$; and almost null systematic errors; Wen et al. 2012) but they are negligible and they cannot be responsible for the present result. The statistical errors almost cancel when we do the stacking (error of the redshift equal to $\sigma _z/\sqrt{N}$).

\subsection{Selection effects}

The incompleteness of the cluster sample is estimated by Wen et al. (2012, Secc. 2.5).
The detection rate is nearly 100\% for clusters with masses $M_{200}>2\times 10^{14}$ M$_\odot $ up to redshifts of $z\approx 0.5$, and $\sim $75\% for clusters of $M_{200}>6\times 10^{13}$ M$_\odot$ up to redshifts of $z\approx 0.42$. In fact, more than 95\% of clusters of $M_{200}>10^{14}$ M$_\odot$ are detected
for $z<0.42$.
This incompleteness should not affect our results because our results do not depend on the counts of clusters and the undetected clusters are not expected to be different
from the detected clusters. The only effect may be a variation of the average mass with redshift, but this is already taken into account in the fits because we
calculate the average for each bin. Indeed, the variation of the average mass
within the five bins A$n$, B$n$ with $n=m,m+5,m+10,m+15,m+20$, with fixed $m=1,2,3,4,5$ (see Tables \ref{Tab:binsa} and \ref{Tab:binsb}), is very small, reflecting that the incompleteness is only slightly varying with redshift at least within $z<0.4$ (see Fig. 6 of Wen et al. 2012).

We check that the incompleteness effects are not important by repeating our
fit of Eq. (\ref{doublefit}), removing the bins with $\langle M_{200}\rangle <10^{14}$ M$_\odot $ or $\langle z\rangle >0.42$, that is, we use only the bins A4, A5, A9, A10, A14, A15; and B2-5, B7-10, B12-15. The result is $\alpha _3=3.23\pm 0.96$, $\alpha _4=3.42\pm 1.61$ for Mode A, and $\alpha _3=1.96\pm 0.27$, 
$\alpha _4=-1.79\pm 2.16$ for Mode B. For other fits, there are not enough points.
These numbers are compatible with those obtained for the whole sample (Table \ref{Tab:fits})
within 1.5$\sigma $, though the error bars here are much larger due to the use of a lower number of points for the fits. Anyway, we still see a self-similar value in the dependence with the mass
($\alpha _3=1.96\pm 0.27$ in Mode B) and we still see a significant evolution with redshift
($\alpha _4=3.42\pm 1.61$ in Mode A), so we can say that the selection effects are not responsible for these trends.

\subsection{Using X-ray luminosity--derived masses from MCXC}

Another possibility  to explain the
present result is that the masses $M_{\rm 200}$ were not correctly calculated and that
an average variation of $\langle M_{\rm 200}\rangle $ with the redshift exists. We cannot attribute
this to the evolution of galaxies, since Wen et al. (2012) have taken it into account.
However, their approach, which is only an extrapolation of the evolution at low $z$, may possibly introduce some bias into their estimation of mass. In particular, at $z\gtrsim  0.42$ some galaxies are not observed and the the catalog is incomplete, so the calculated richness will contain some bias.

\begin{figure}
\vspace{.5cm}
\centering
\includegraphics[width=7.2cm]{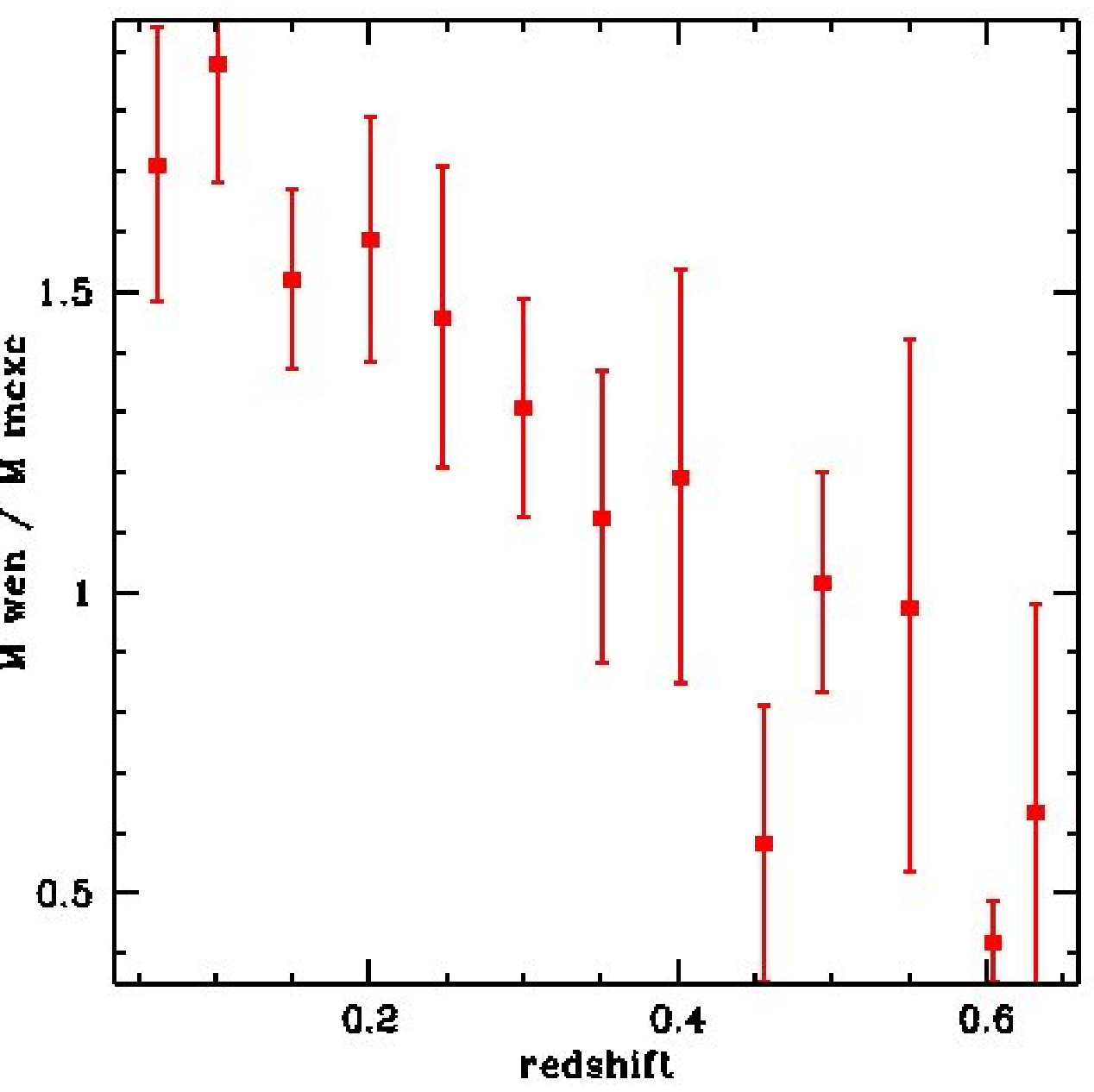}
\\ \vspace{1cm} 
\includegraphics[width=7.2cm]{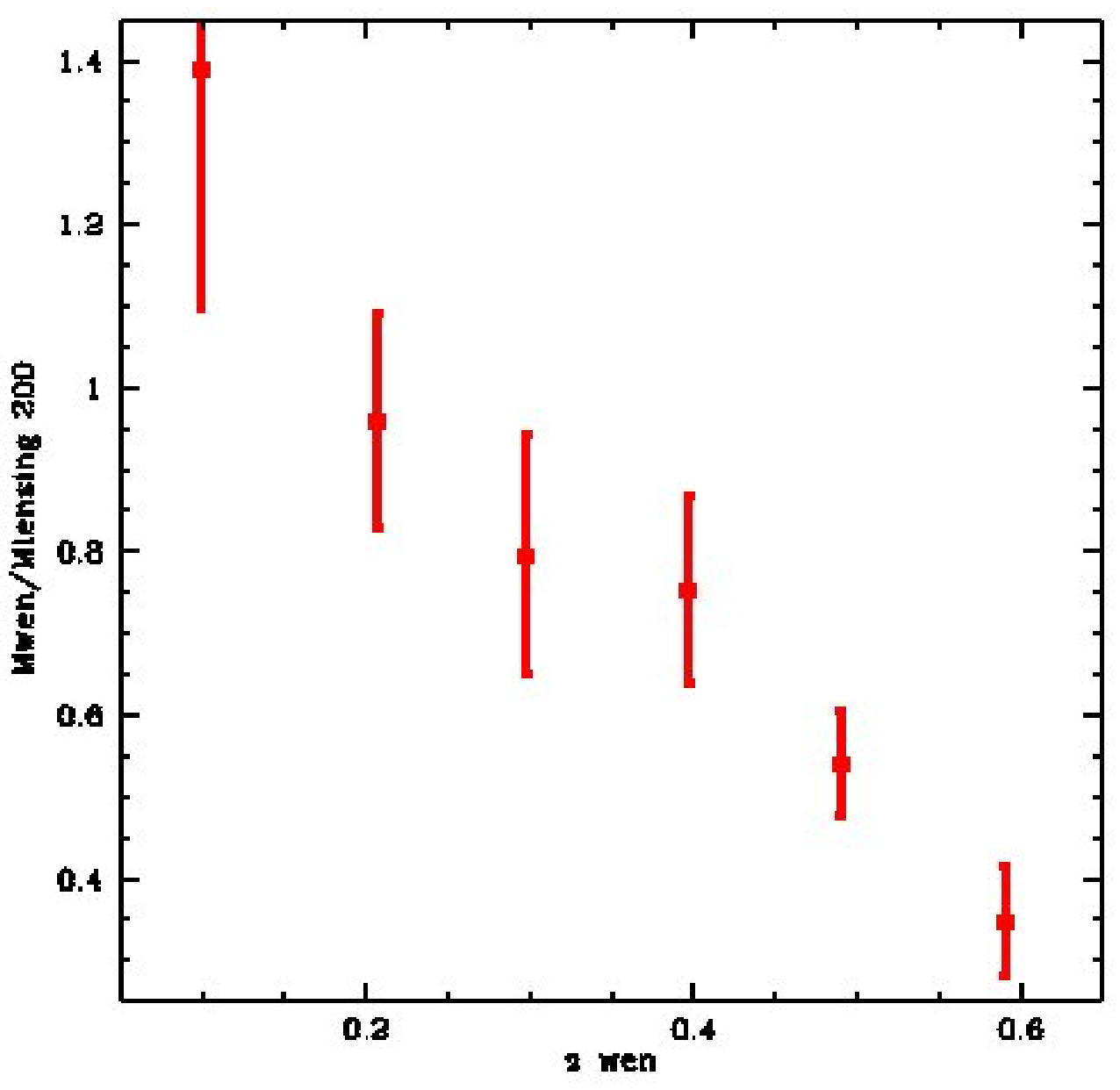}
\caption{Comparison of the average of the ratio of $\frac{M_{\rm 200}}{M_{\rm 500,MCXC}}$ (top)
and $\frac{M_{\rm 200}}{M_{\rm 200,{\rm weak\ lensing}}}$ (bottom)
for different bins of redshift, using a total of 414 or 194 clusters respectively common in MCXC or weak lensing and Wen et al. (2012) catalogue.}
\label{Fig:comparamasas}
\end{figure}

In order to explore the hypothesis of this miscalculation of masses in Wen et al., we have cross-correlated the $M_{\rm 200}$ masses given by those authors with the masses of 414 common objects (coincident position within 1 arcminute) of the catalog MCXC of clusters detected
by Planck and with masses $M_{\rm 500,MCXC}$ derived from X-ray
(Piffaretti et al. 2011). Figure \ref{Fig:comparamasas} (top) gives the dependence of the ratio of 
$\frac{M_{\rm 200}}{M_{\rm 500,MCXC}}$ with redshift. There is also some dependence
on mass. On average, using again a fit of the type 
$\frac{M_{\rm 200}}{M_{\rm 500,MCXC}}=R_5\,\left(\frac{M_{\rm 200}}{10^{14}\,M_\odot}\right)^{\alpha _5}(1+z)^{\alpha _6}$, 
we get $R_5=1.39\pm 0.11$, $\alpha _5=0.34\pm 0.04$, and $\alpha _6=-1.72\pm 0.27$.
These values of $\alpha _5$ and $\alpha _6$ may change slightly if we select clusters
within some limited range of redshift or masses 
instead of the whole sample of 414 clusters, but 
the results are anyway similar: there is no significant variation of
these exponents with redshift and the variation with mass is more conspicuous, but compatible
with the global analysis within the errors.
It is a very significant departure from an almost constant $\frac{M_{\rm 200}}{M_{\rm 500,MCXC}}$ that would be expected theoretically if the estimations of both masses were correct: $\approx 1.40(1+z)^{0.088}$ for Navarro--Frenk--White (NFW) density distribution, or 1.58 for a singular isothermal distribution (see appendix \ref{.ratio200_500}). 
Taking NFW, we see that the average correction to apply to $M_{200}$ given by Wen et al.  
in order to be compatible with the values of $M_{500}$ of MCXC is

\[
M_{200}^{corr.}=(1.01\pm 0.08) M_{200}
\left(\frac{M_{\rm 200}}{10^{14}\,M_\odot}\right)^{-0.34\pm 0.04}
\]\begin{equation}\times
(1+z)^{1.81\pm 0.27}
\label{m200corr}
\end{equation}

\begin{figure}
\vspace{.5cm}
\centering
\includegraphics[width=8cm]{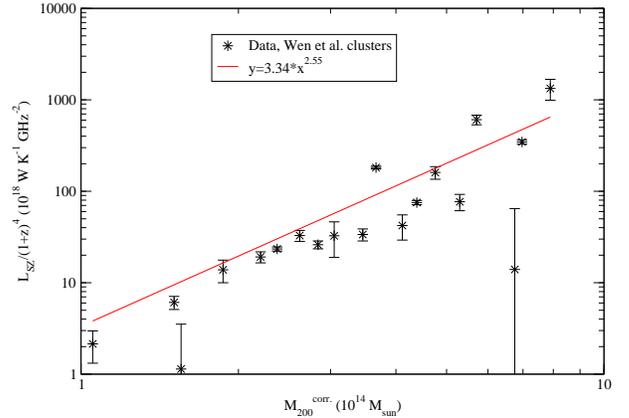}
\caption{Luminosity-SZ vs. $M_{200}^{corr.}$ (mass corrected assuming MCXC values)
for Mode B. Error bars stand only for the error in the luminosity for each bin.}
\label{Fig:X_LvsM}
\end{figure}

With this correction, we can calculate again the parameters $\alpha _i$ of the best
fit, substituting $M_{200}$ for $M_{200}^{corr.}$. The new results are given in
Table \ref{Tab:fitscorr}. Here we have not grouped the bins of the same mass because, with the
correction, the dispersion of values is higher.
The systematic errors of Eq. (\ref{m200corr}) are, however, considered in the error bars of the derived parameters; the statistical errors for each bin of clusters
are considered negligible given the large number of clusters per bin. 
The result is again problematic but in a different sense: now we see that dependence of the luminosity with mass has an exponent larger than 5/3, between 2.2 and 2.7 
(maximum significance of the difference: 2.8$\sigma $ in $\alpha _3$ for Mode B); this is illustrated
in Fig. \ref{Fig:X_LvsM}. Furthermore, 
either there is no evolution of luminosity with redshift (for mode A) or there is an 
evolution with redshift opposite to what was observed previously 
($\alpha _4<0$ for mode B). 
We solve with it a result that was not understood with the masses
of Wen et al. (2012) but we get a new problematic result:
the self-similar value of the dependence of the luminosity mass with the mass is not
obtained, which might be due to the fact that the clusters are not virialized and in the conditions of the simulations that obtain the exponent of 5/3. This is something that
was already questioned by Czakon et al. (2015), but they relate it to possible biases.
However, again, we may doubt about the validity of X-ray masses of MXCX catalog.
The method to derive these masses may introduce some bias depending on the
mass (von der Linden et al. 2014, Hoekstra et al. 2015).

At least one of the two mass estimators must be wrong, and it is possible that both are wrong. It cannot be a geometric or morphological question in the cluster given by the different radii up to which the mass or luminosity is integrated: we show in appendix \ref{.ratio200_500} that the difference between $M_{200}$ and $M_{500}$ should be a constant factor, almost independent of redshift. Our guess is that the evolution correction measured by Wen et al. (2012) in order to convert richness into mass is not accurate enough, and it may be possibly affected by some bias of incompleteness for clusters with $z>0.42$, which do not allow us to count all the galaxies embedded in a cluster and thus produce an error in the estimation of the richness. In a previous work, Wen et al. (2009) claimed that their mass
determinations were compatible with X-ray measurements, but these comparisons were
limited to redshifts $z<0.26$, and, moreover, they used a different definition of richness. It is at higher redshift where we find significant differences between the masses of Wen et al. (2012) and MCXC masses.
The analysis done by Wen \& Han (2015) has also found that there is a difference between Wen et al. (2012) masses and the X-ray masses, which may be corrected with a better estimation of
the evolution factor in the cluster richness--luminosity relationship.
On the other hand, for X-ray measurements of MCXC masses, we may suspect that the method of Piffaretti et al. (2011) assuming a particular mass-luminosity ratio may be biased or that the extrapolation of the cluster mass profile to its
outer parts may contain some inaccuracy. Piffaretti et al. (2011) also warn that the $M_{500,MCXC}$ values provided by the MCXC rely on the assumption that, on average, the Malmquist bias for the samples used to construct the MCXC is the same as that of the REXCESS sample.
The X-ray luminosities are also affected by two main sources of uncertainty (Piffaretti et al. 2011): uncertainties in the underlying model used in the homogenization procedure and the measurement uncertainty in the original input catalog.

\begin{table}
\caption{Parameters of $L_{\rm SZ}$ (units: $10^{18}$ W K$^{-1}$ GHz$^{-2}$) in the dependence on the corrected mass $M_{200}^{corr.}$ (see Eq. \ref{m200corr}) and redshift
for bins of mode A and mode B. Errors include the errors of the fits and the uncertainties
in the parameters of Eq. (\ref{m200corr}).}
\begin{center}
\begin{tabular}{ccc}
Parameter & Mode A & Mode B  \\ \hline
$L_1$ & $3.22\pm 1.29$ & $3.34\pm 1.69$ \\
$\alpha _1$ & $2.20\pm 0.45$ & $2.55\pm 0.36$ \\
$L_2$  & $1.78\pm 2.64$ & $12.0\pm 10.2$ \\ 
$\alpha _2$ & $7.24\pm 3.91$ & $1.83\pm 2.12$ \\
$L_3$ & $2.87\pm 1.31$ & $5.01\pm 2.61$\\ 
$\alpha _3$ & $2.51\pm 0.55$ & $2.70\pm 0.37$ \\
$\alpha _4$ & $-0.78\pm 1.12$ & $-3.39\pm 1.52$ \\ \hline
\end{tabular}
\end{center}
\label{Tab:fitscorr}
\end{table}

\subsection{Using weak-lensing derived masses}

Weak-lensing masses of clusters (von der Linden et al. 2014; Applegate et al. 2014; Hoekstra et al. 2015; Sereno 2015; Melchior et al. 2016) are less affected by the bias of the present optical 
or X-ray derived masses, and they will be useful to solve this question. 

We cross-correlate the $M_{\rm 200}$ masses given by Wen et al. (2012) with the weak-lensing masses of clusters in the compilation by Sereno (2015): there are 194 objects with data in both samples. 
On average, the fit of type
$\frac{M_{\rm 200}}{M_{\rm 200,lensing}}=R_5\,\left(\frac{M_{\rm 200}}{10^{14}\,M_\odot}\right)^{\alpha _5}(1+z)^{\alpha _6}$ gives: $R_5=0.68\pm 0.13$, $\alpha _5=0.29\pm 0.06$, and $\alpha _6=-1.84\pm 0.51$.
Fig. \ref{Fig:comparamasas} (bottom) gives the dependence of the ratio of 
$\frac{M_{\rm 200}}{M_{\rm 200,lensing}}$ with redshift.
Therefore, the average correction to apply to $M_{200}$ given by Wen et al.  
in order to be compatible with these values of weak lensing is

\[
M_{200}^{corr. 2}=(1.45\pm 0.27) M_{200}
\left(\frac{M_{\rm 200}}{10^{14}\,M_\odot}\right)^{-0.29\pm 0.06}
\]\begin{equation}\times 
(1+z)^{1.84\pm 0.51}
\label{m200corr2}
\end{equation}

This correction is very similar to that of Eq. (\ref{m200corr}) using X-ray, except only for
the amplitude, which lead us to think that X-ray derived masses are not so biased.
With this new correction, we can calculate again the parameters $\alpha _i$ of the best
fit substituting $M_{200}$ for $M_{200}^{corr. 2}$. The new results are given in
Table \ref{Tab:fitscorr2}. Again, we see no significant evolution but some departure
from the self-similar exponent of 5/3 in the mass dependence, though less significant
in this case (maximum significance of the difference: 2.2$\sigma $ in $\alpha _3$ for Mode B):
a value of $\alpha _1$ or $\alpha _3$ between 2.1 and 2.5.

\begin{table}
\caption{Parameters of $L_{\rm SZ}$ (units: $10^{18}$ W K$^{-1}$ GHz$^{-2}$) in the dependence on the corrected mass $M_{200}^{corr. 2}$ (see Eq. \ref{m200corr2}) and redshift 
for bins of mode A and mode B. Errors include the errors of the fits and the uncertainties
in the parameters of Eq. (\ref{m200corr2}).}
\begin{center}
\begin{tabular}{ccc}
Parameter & Mode A & Mode B  \\ \hline
$L_1$ & $1.42\pm 1.02$ & $1.42\pm 1.04$ \\
$\alpha _1$ & $2.14\pm 0.53$ & $2.40\pm 0.36$ \\
$L_2$  & $1.78\pm 2.64$ & $12.0\pm 10.2$ \\ 
$\alpha _2$ & $7.24\pm 3.91$ & $1.83\pm 2.12$ \\
$L_3$ & $1.23\pm 0.90$ & $2.02\pm 1.57$\\ 
$\alpha _3$ & $2.34\pm 0.54$ & $2.51\pm 0.38$ \\
$\alpha _4$ & $-0.53\pm 2.04$ & $-3.12\pm 1.89$ \\ \hline
\end{tabular}
\end{center}
\label{Tab:fitscorr2}
\end{table}

In this case, we cannot think about any possible systematic bias, and the fact of the coincidence with
the result of X-ray measurements of masses corroborates this scenario as the most likely one:
no (significant) evolution and $L_{SZ}\propto \sim M^3$.

\subsection{Number and temperature of electrons}

Using the value of $L_i$ with higher signal-to-noise, $L_1=(3.93\pm 0.77)\times 10^{18}$
W K$^{-1}$ GHz$^{-2}$ for mode B analysis, the equation (\ref{lsz}) allows us to estimate
the average redshift of the sample
 and that of cluster masses of $M_{200}=10^{14}$ M$_\odot $ for which
$N_e\langle T_e(K)\rangle =(9.07\pm 1.78)\times 10^{76}$ per cluster.
An electronic temperature of $T_e\sim 2\times 10^7$ K (Vikhlinin et al.
2006) gives around $5\times 10^{69}$ electrons per cluster. With $N_e=f_eN_H$,
where $f_e$ is the number of electrons per hydrogen atom of the cluster gas and $N_H$ is
the number of hydrogen atoms, and estimating the mass of the gas as $M_{\rm gas}=N_H\,m_{\rm proton } \left(1+4\frac{Y}{1-Y}\right)$ ($Y$ is the helium abundance; we assume it is equal to 0.25; we neglect the weight of the atoms heavier than helium), we get
$M_{\rm gas}\sim 10^{13}/f_e$ M$_\odot $, that is, $\frac{M_{\rm gas}}{M_{200}}\sim 0.10\,f_e^{-1}$.
Roughly, $f_e\sim 1$, because most atoms are hydrogen and are ionized, so this means that the gas mass is a 10\% of the total mass, as previously observed (Vikhlinin et al. 2006).

\section{Summary and conclusions}

We have used the PLANCK-DR2 data and a catalog 
with $\sim 10^5$ clusters of galaxies
to measure its SZ flux variation once an average dust emission from clusters was subtracted.
We analyzed the dependence of the amplitude of this effect 
versus the redshift and the mass of the clusters.
The absolute value of the amplitude indicates that
the gas mass is around 10\% of the total mass (for cluster masses 
of $\sim 10^{14}$ M$_\odot $). 
The dependence of this amplitude with the mass or with the redshift appears
to contradict the expectations.
Either we find an unexpected 
evolution with redshift of this amplitude for fixed mass (proportional to $(1+z)^{3.9\pm 0.8}$, in mode A binning) if we take the masses measured from visible catalogs by Wen et al. (2012) and take into account the evolution factors, or there is a significant 
departure of the self-similar dependence
of the amplitude (proportional to $M^{2.5-2.7}$ instead of proportional 
to $M^{5/3}$, in mode B binning) if we take the corrected masses to
fit measured values using X-ray values or weak lensing.
Given that the last two methods give the same results and that at least weak lensing 
is a priori expected to be free of important biases, we conclude that the second result is more likely: that cluster masses of Wen et al. (2012) are underestimated at high redshift
and the amplitude of the luminosity in the SZ scales with the mass as $L_{\rm SZ}\propto M^B$ with an index $B=2.70\pm 0.37$ 
(using X-ray derived masses) or 
$B=2.51\pm 0.38$ (using weak-lensing derived masses) for a range of cluster masses (X-ray or weak lensing) between $1\times 10^{14}$ M$_\odot $ and 
$8\times 10^{14}$ M$_\odot $.

\begin{acknowledgements}
Thanks are given to the anonymous referee for helpful suggestions that improved this paper.
Funding for SDSS-III has been provided by the Alfred P. Sloan Foundation, the Participating Institutions, the National Science Foundation, and the U.S. Department of Energy Office of Science. The SDSS-III web site is http://www.sdss3.org/.

SDSS-III is managed by the Astrophysical Research Consortium for the Participating Institutions of the SDSS-III Collaboration including the University of Arizona, the Brazilian Participation Group, Brookhaven National Laboratory, Carnegie Mellon University, University of Florida, the French Participation Group, the German Participation Group, Harvard University, the Instituto de Astrofisica de Canarias, the Michigan State/Notre Dame/JINA Participation Group, Johns Hopkins University, Lawrence Berkeley National Laboratory, Max Planck Institute for Astrophysics, Max Planck Institute for Extraterrestrial Physics, New Mexico State University, New York University, Ohio State University, Pennsylvania State University, University of Portsmouth, Princeton University, the Spanish Participation Group, University of Tokyo, University of Utah, Vanderbilt University, University of Virginia, University of Washington, and Yale University. 

Some of the results
in this paper have been derived using the
HEALPix package. The research leading to the PLANCK results has received funding from the European Research Council under   the   European   Union’s   Seventh   Framework   Programme   (FP7/2007-2013)/ERC  grant  agreement  No.  307209,  as  well  as  funding  from  an  STFC Consolidated  Grant  (No.  ST/L000768/1).  The  development  of
Planck has been  supported  by:  ESA;  CNES  and  CNRS/
INSU-IN2P3-INP  (France);  ASI,
CNR,  and  INAF  (Italy);  NASA  and  DoE  (USA);  STFC  and  UKSA  (UK);
CSIC,  MICINN,  JA,  and  RES  (Spain);  Tekes,  AoF,  and  CSC  (Finland);
DLR and MPG (Germany); CSA (Canada); DTU Space (Denmark); SER/SSO (Switzerland);  RCN  (Norway);  SFI  (Ireland);  FCT/MCTES  (Portugal);  and
PRACE  (EU).  A  description  of  the  Planck  Collaboration  and  a  list  of  
its members,  including  the  technical  or  scientific  activities  in  which  they  have been involved, can be found at
http://www.sciops.esa.int/index.php?
project=planck\&page=Planck\_Collaboration

\end{acknowledgements}

\appendix

\section{Corrections of the flux measurement}

\subsection{Method to correct superposition of several clusters in a line of sight} 
\label{.sevclu}

\begin{figure}
\vspace{.5cm}
\centering
\includegraphics[width=8cm]{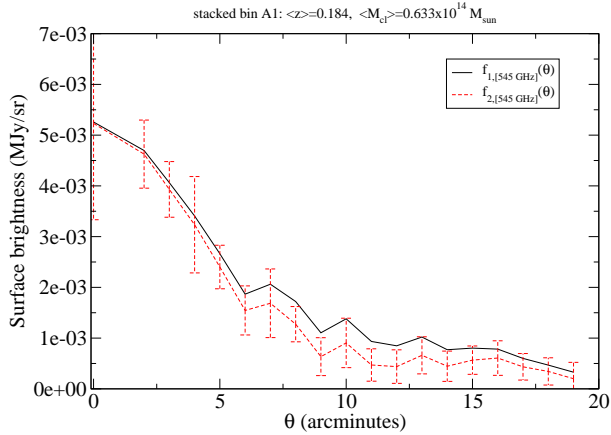}
\caption{Average measured surface brightness per cluster as a function of the distance to the center of the cluster, with sky emission subtracted, for bin A1 at 545 GHz before correction
by cluster correlation ($f_1$) and after correction by cluster correlation ($f_2$).}
\label{Fig:bin1_corrcorr}
\end{figure}

Several clusters may be observed in a given line of sight.
One way to take into account this and calculate the emission per cluster is 
\begin{equation}
f_{2,\nu}(\theta )=f_{1,\nu}(\theta )-\frac{1}{2\pi }\int _0^{2\pi }d\phi ' \int _0^\infty d\theta '\,\theta '
 C(\theta ')\times f_{1,\nu}\left(\sqrt{\theta ^2+\theta '^2-2\,\theta \,\theta '\,
\cos(\phi ')}\right)
,\end{equation}\[
C(\theta )=\sqrt{\langle \sigma \sigma \rangle (\theta )-\langle \sigma \rangle ^2}
,\]
where $\sigma $ is the surface density of clusters.
These amounts are obtained from the analysis of our
sample. 
An example of the correction
is given in Fig. \ref{Fig:bin1_corrcorr} for the stacked bin A1 in the frequency of 545 GHz.
For small correlations, as it is our case, one iteration is enough; otherwise, we could get a better approximation by calculating in the same
way $f_{3,\nu}(\theta )$, $f_{4,\nu}(\theta )$,...

Finally, we derive the integrated flux of the cluster within $\theta _{\rm max}$ by means of:
\begin{equation}
\label{fluxint}
F_{\nu, \theta _{\rm max}}=2\,\pi\,\int _0^{\theta _{\rm max}}d\theta \,\theta \,f_{2,\nu}(\theta )
.\end{equation}

The error bars are calculated with the propagation of errors of all the processes.

\subsection{Correction of flux lost due to beam dilution}
\label{.corrfluxloss}

There is some small amount of flux that we loose for angular distances larger than 
$\theta _{\rm max}$. This is stronger for low frequencies, where the telescope beam profile is wider.
Assuming a Gaussian profile, their corresponding sigmas are 5.67' for 70 GHz (Planck collaboration 2016b); 4.12' for 100 GHz, 3.11' for 143 GHz, 2.14' for 217 GHz,
2.10' for 353 GHz, 2.06' for 545 GHz, 1.97' for 857 GHz (Planck collaboration 2016c); and 0.85' for 2998 GHz (Miville-Deschenes \& Lagache  2005). The closest and most massive clusters may have an angular size of 
$R_{200}$ comparable to $\theta _{\rm max}$: for instance, at $z=0.12$, 10.5 arcminutes correspond to
1.4 Mpc (for Abell clusters, the angular size is larger, 
but they will not be used in the statistics of the amplitude of the flux).
Planck beam is very close to Gaussian, especially for frequencies below 545 GHz. Moreover, stacking of clusters produce profiles with high circular symmetry, reducing fluctuations. However, there may be departures from Gaussianity due to the own shape of the clusters and the sum of clusters with different angular size in the stacking, due to the dispersion of redshifts and masses in each bin; we neglect them since this correction of flux is of second order.

Numerical experiments on the integration of the flux showed us that increasing the value
of the maximum radius of integration with $\theta _{\rm max}>10.5'$ does not improve the result, because the tail of the Gaussian introduces mostly noise. The fit to a Gaussian
would be a solution, but it reduces the accuracy with respect to our method of integration.
Instead, we do the following approach of extrapolation: assuming Gaussianity in $f_{2,\nu}$ with rms $\sigma $, we get from Eq. (\ref{fluxint}) $F_{\nu, \theta _{\rm max}}\propto
\sigma ^2\,\left[1-exp\left( -\frac{\theta _{\rm max}^2}{2\sigma ^2}\right)\right]$
and consequently
\begin{equation}
F_\nu\equiv F_{\nu, \infty}=F_{\nu, \theta _{max}}\times {\rm factor}
,\end{equation}
\[
{\rm factor}=\frac{1}{1-y};\ \ \ 
y\equiv exp\left( -\frac{\theta _{\rm max}^2}{2\sigma ^2}\right)
.\]
Using another angle $\theta _0=\frac{\theta _{\rm max}}{\sqrt{2}}\approx 7.5'$, we can
derive $y$ from the ratio $r=\frac{F_{\nu, \theta _{max}}}{F_{\nu, \theta _0}}$:
\begin{equation}
r=\frac{1-y}{1-\sqrt{y}}
,\end{equation}
thus,
\begin{equation}
F_\nu=F_{\nu, \theta _{max}}\frac{1}{1-\left(\frac{F_{\nu, \theta _{max}}}{F_{\nu, \theta _0}}-1\right)^2}
.\end{equation}

This correction is relatively small, $\lesssim 10$\%, being the most significant for
70 GHz.

\section{Tables of results}

Tables \ref{Tab:binsa} and \ref{Tab:binsb} give the numbers obtained of the SZ analysis.

\begin{table*}
\scriptsize
\caption{Stacked bins of clusters, mode A, and the parameters obtained in their fits. The number of clusters in each bin is $N_{\rm cl}$.}
\begin{center}
\begin{tabular}{|cccc|cc|cc|cc|}
\#  & $N_{\rm cl}$ & $\langle z\rangle$ & $\left\langle \frac{M_{\rm 200}}{10^{14}\,M_\odot}
\right\rangle $ & 
$\beta _{\rm dust}$ & $T_{\rm dust}$ (K) &
$A_{\rm SZ}$ ($\mu $Jy K$^{-1}$ GHz$^{-2}$) & $T_{\rm CMB}$ (K) & 
$\frac{L_{\rm SZ}}{(1+z)^4}$ ($10^{18}$ W K$^{-1}$ GHz$^{-2}$) & $\frac{F_{\rm SZ}}{F_{\rm dust}}$(545 GHz) \\ \hline
A1 &    2710 &  0.184  & 0.633 & 
1.52 & 25 &
$0.036\pm 0.022$ & $3.41^{+2.51}_{-2.21}$ & $1.7\pm 1.1$ & 0.029 \\
A2 &    3359 &  0.181  & 0.742 & 
1.24 & 27 &
$0.130\pm 0.085$ & $2.50^{+3.40}_{-1.31}$ & $6.2\pm 4.1$ & 0.014 \\
A3 &    3604 &  0.180  & 0.912 & 
2.44 & 22 &
$0.055\pm 0.020$ & $3.19^{+0.94}_{-0.64}$ & $2.6\pm 1.0$ & 0.128 \\
A4 &    3148 &  0.180  & 1.150 & 
1.60 & 23 &
$0.012\pm 0.006$ & $5.10^{+0.80}_{-3.91}$ & $0.6\pm 0.3$ & 0.053 \\
A5 &    4215 &  0.176  & 2.393 & 
1.84 & 23 &
$0.541\pm 0.020$ & $3.21^{+0.14}_{-0.13}$ & $24.6\pm 0.9$ & 0.233 \\
A6 &    2930 &  0.293  & 0.632 & 
1.28 & 24 &
$-0.372\pm 0.161$ & --- & $-36.5\pm 15.8$ & -0.009 \\
A7 &    3651 &  0.294  & 0.738 & 
1.68 & 37 &
$0.003\pm 0.005$ & $4.44^{+2.03}_{-3.13}$ & $0.3\pm 0.4$ & 0.023 \\ 
A8 &    3805 &  0.294  & 0.912 & 
1.28 & 26 &
$0.071\pm 0.031$ & $3.50^{+1.83}_{-1.00}$ & $7.0\pm 3.1$ & 0.077 \\
A9 &    3089 &  0.293  & 1.152 & 
1.80 & 25 &
$0.070\pm 0.025$ & $3.66^{+2.81}_{-2.35}$ & $6.9\pm 2.4$ & 0.121 \\
A10 &   3556 &  0.291  & 2.252 & 
1.96 & 22 &
$0.251\pm 0.019$ & $3.49^{+0.22}_{-0.19}$ & $24.4\pm 1.9$ & 0.168 \\
A11 &   3104 &  0.382  & 0.632 & 
2.16 & 23 &
$-0.067\pm 0.212$ & --- & $-9.4\pm 30.0$ & 0.000 \\ 
A12 &   3786 &  0.381  & 0.739 & 
1.60 & 30 &
$-0.486\pm 0.305$ & --- & $-67.5\pm 42.4$ & 0.000 \\
A13 &   3873 &  0.380  & 0.911 & 
1.44 & 28 &
$0.030\pm 0.020$ & $4.14^{+2.76}_{-1.66}$ & $4.2\pm 2.8$ & 0.021 \\
A14 &   3130 &  0.380  & 1.149 & 
1.44 & 27 &
$0.023\pm 0.024$ & $4.47^{+2.43}_{-3.08}$ & $3.2\pm 3.3$ & 0.041 \\
A15 &   3093 &  0.378  & 2.102 & 
2.00 & 25 &
$0.218\pm 0.023$ & $3.67^{+0.41}_{-0.35}$ & $30.0\pm 3.2$ & 0.094 \\   
A16 &   3056 &  0.459  & 0.633 & 
4.00 & 12 &
$0.060\pm 0.124$ & $2.71^{+4.58}_{-1.14}$ & $10.4\pm 21.5$ & 0.007 \\
A17 &   4009 &  0.461  & 0.740 & 
1.52 & 21 &
$-0.337\pm 0.206$ & --- & $26.8\pm 12.8$ & $-58.7$ \\
A18 &   4129 &  0.460  & 0.912 & 
1.36 & 34 &
$-0.294\pm 0.313$ & --- & $-51.2\pm 54.5$ & 0.000 \\    
A19 &   3052 &  0.460  & 1.146 & 
2.20 & 23 &
$0.140\pm 0.035$ & $3.61^{+0.76}_{-0.53}$ & $24.3\pm 6.1$ & 0.054 \\  
A20 &   2822 &  0.460  & 2.006 & 
2.23 & 23 &
$0.209\pm 0.028$ & $3.80^{+0.39}_{-0.34}$ & $36.3\pm 4.8$ & 0.126 \\   
A21 &   2565 &  0.568  & 0.634 & 
3.88 & 11 &
$-0.442\pm 0.266$ & --- & $-95.9\pm 57.7$ & -0.002 \\
A22 &   3807 &  0.572  & 0.742 & 
1.44 & 31 &
$0.041\pm 0.023$ & $4.83^{+3.03}_{-1.62}$ & $9.0\pm 5.0$ & 0.044 \\ 
A23 &   4472 &  0.579  & 0.914 & 
1.76 & 24 &
$0.019\pm 0.022$ & $4.01^{+3.88}_{-2.42}$ & $4.2\pm 4.9$ & 0.008 \\  
A24 &   3519 &  0.583  & 1.150 &
2.84 & 21 & 
$0.224\pm 0.031$ & $3.66^{+0.59}_{-0.43}$ & $49.8\pm 9.2$ & 0.105 \\   
A25 &   2676 &  0.580  & 1.921 &
2.16 & 23 &
$0.198\pm 0.023$ & $4.35^{+0.41}_{-0.33}$ & $43.8\pm 5.0$ & 0.192 \\ \hline
A26 &   394  &  0.0526 & --- & 
1.52 & 24 &
$3.33\pm 0.08$ & $2.88^{+0.21}_{-0.17}$ & $18.2\pm 0.4$ & 0.797 \\
A27 &   395  &  0.152 & --- &
1.52 & 24 &
$2.98\pm 0.05$ & $3.11^{+0.21}_{-0.17}$ & $106.4\pm 2.1$ & 0.344 \\ \hline
\end{tabular}
\end{center}
\label{Tab:binsa}
\end{table*}

\begin{table*}
\scriptsize
\caption{Stacked bins of clusters, mode B, and the parameters obtained in their fits.
The number of clusters in each bin is $N_{\rm cl}$.}
\begin{center}
\begin{tabular}{|cccc|cc|cc|cc|}
\#  & $N_{\rm cl}$ & $\langle z\rangle$ & $\left\langle \frac{M_{\rm 200}}{10^{14}\,M_\odot}
\right\rangle $ & 
$\beta _{\rm dust}$ & $T_{\rm dust}$ (K) &
$A_{\rm SZ}$ ($\mu $Jy K$^{-1}$ GHz$^{-2}$) & $T_{\rm CMB}$ (K) & 
$\frac{L_{\rm SZ}}{(1+z)^4}$ ($10^{18}$ W K$^{-1}$ GHz$^{-2}$) & $\frac{F_{\rm SZ}}{F_{\rm dust}}$(545 GHz) \\ \hline
B1 &    2582 &  0.125  & 0.771 & 
1.32 & 25 &
$0.084\pm 0.032$ & $3.51^{+1.95}_{-1.01}$ & $2.4\pm 0.7$ & 0.061 \\
B2 &    1818 &  0.123  & 1.334 & 
1.36 & 24 &
$0.244\pm 0.040$ & $3.06^{+0.47}_{-0.37}$ & $4.5\pm 0.6$ & 0.066 \\
B3 &    510 &  0.121  & 2.664 & 
2.52 & 19 &
$0.964\pm 0.049$ & $2.94^{+0.25}_{-0.21}$ & $15.7\pm 0.9$ & 0.199 \\
B4 &    133 &  0.119  & 5.178 & 
3.40 & 18 &
$7.71\pm 0.16$ & $3.19^{+0.07}_{-0.07}$ & $97.8\pm 3.4$ & 6.22 \\
B5 &    15 &  0.115  & 10.231 & 
1.24 & 57 &
$27.3\pm 2.6$ & $3.18^{+2.40}_{-2.05}$ & $255.5\pm 21.1$ & -37.3 \\
B6 &    14249 &  0.246  & 0.761 & 
1.52 & 27 &
$-0.151\pm 0.264$ & --- & $-11.2\pm 20.0$ & 0.000 \\
B7 &    8264 &  0.244  & 1.312 & 
1.64 & 25 &
$-1.26\pm 2.5$ & --- & $5.8\pm 0.8$ & 0.000 \\ 
B8 &    2006 &  0.239  & 2.660 & 
1.24 & 27 &
$0.356\pm 0.035$ & $3.33^{+0.30}_{-0.25}$ & $23.8\pm 2.3$ & 0.171 \\
B9 &    413 &  0.239  & 5.141 & 
1.80 & 23 &
$1.03\pm 0.05$ & $3.37^{+0.21}_{-0.19}$ & $67.5\pm 3.8$ & 0.190 \\
B10 &   51 &  0.236  & 10.445 & 
1.44 & 25 &
$4.83\pm 0.14$ & $3.39^{+0.14}_{-0.11}$ & $316.4\pm 16.3$ & 0.572 \\
B11 &   20667 &  0.401  & 0.762 & 
1.36 & 29 &
$0.008\pm 0.016$ & $4.25^{+2.76}_{-2.83}$ & $8.05\pm 5.5$ & 0.011 \\ 
B12 &  10828  &  0.399  & 1.300 & 
1.88 & 25 &
$0.130\pm 0.018$ & $3.58^{+0.49}_{-0.39}$ & $14.5\pm 2.1$ & 0.067 \\
B13 &   1925 &  0.396  & 2.585 & 
2.00 & 24 &
$0.231\pm 0.035$ & $3.74^{+0.50}_{-0.39}$ & $29.3\pm 4.5$ & 0.082 \\
B14 &   242 &  0.386  & 5.035 & 
2.32 & 23 &
$0.544\pm 0.110$ & $3.69^{+0.78}_{-0.58}$ & $75.0\pm 15.1$ & 0.404 \\
B15 &   12 &  0.378  & 9.360 & 
2.00 & 15 &
$9.68\pm 2.46$ & $3.39^{+0.90}_{-0.66}$ & $562.8\pm 191.0$ & 0.269 \\   
B16 &   11818 &  0.544  & 0.772 &
1.84 & 27 &
$0.067\pm 0.019$ & $3.97^{+1.56}_{-0.91}$ & $12.5\pm 3.4$ & 0.034 \\
B17 &   6224 &  0.547  & 1.281 & 
3.04 & 20 &
$0.157\pm 0.022$ & $3.85^{+1.36}_{-0.79}$ & $29.9\pm 4.1$ & 0.200 \\
B18 &   836 &  0.539  & 2.576 & 
1.36 & 28 &
$0.205\pm 0.063$ & $4.40^{+1.66}_{-0.94}$ & $41.4\pm 12.7$ & 0.139 \\    
B19 &   69 &  0.531  & 4.765 & 
4.00 & 13 &
$-1.55\pm 1.62$ & --- & $11.3\pm 10.9$ & 0.00 \\  
B20 &   2 &  0.526  & 9.388 & 
4.00 & 25 &
$-1.75\pm 9.72$ & --- & $-335.4\pm 1900.2$ & -25.7 \\   
B21 &   1331 &  0.671  & 0.800 & 
2.32 & 16 &
$-0.207\pm 0.255$ & --- & $-76.9\pm 96.1$ & 0.000 \\
B22 &   1061 &  0.679  & 1.288 & 
1.96 & 22 &
$0.128\pm 0.054$ & $3.93^{+2.23}_{-1.02}$ & $31.1\pm 13.0$ & 0.014 \\ 
B23 &   98 &  0.681  & 2.519 & 
2.36 & 24 &
$0.639\pm 0.099$ & $4.74^{+3.61}_{-3.04}$ & $123.3\pm 19.2$ & 0.165 \\  
B24 &   6 &  0.671  & 4.352 & 
4.00 & 23 &
$0.055\pm 0.201$ & $8.36^{+0.17}_{-6.67}$ & $57.5\pm 76.7$ & 0.214 \\   
B25 &   0 &  ---  & --- &
--- & --- &
 --- & --- & --- & --- \\ \hline
B26 &   394  &  0.0526 & --- & 
1.52 & 24 &
$3.33\pm 0.08$ & $2.88^{+0.21}_{-0.17}$ & $18.2\pm 0.4$ & 0.797 \\
B27 &   395  &  0.152 & --- & 
1.52 & 24 &
$2.98\pm 0.05$ & $3.11^{+0.21}_{-0.17}$ & $106.4\pm 2.1$ & 0.344 \\ \hline
\end{tabular}
\end{center}
\label{Tab:binsb}
\end{table*}

\section{Ratio of $M_{200}/M_{500}$}
\label{.ratio200_500}

We will use two models of distribution of $\rho (r)$ in clusters of galaxies to
derive the value of $M_{200}/M_{500}$:

\begin{enumerate}
\item Singular isothermal sphere:
\begin{equation}
\rho (r)=\frac{A}{r^2}
,\end{equation}
so the mass within a radius $R_x$ is
\begin{equation}
M_x\equiv M(R_x)=4\pi \int _0^{R_x}dr\,r^2\rho(r)=4\pi \,A\,R_x
,\end{equation}\begin{equation}
M_x=\frac{4}{3}\pi\,R_x^3\,x\,\rho_c(z)  
,\end{equation}
where $\rho _c(z)=\frac{3H(z)^2}{8\pi G}$ is the critical density at redshift $z$,
$H(z)$ is the corresponding Hubble parameter,
and $x$ is the ratio between the mean density of the cluster up to distance $R_x$ from the center and the critical density. Hence,
\begin{equation}
R_x=\sqrt{\frac{3A}{x\,\rho _c(z)}}
\end{equation}
and, therefore
\begin{equation}
\frac{R_{200}}{R_{500}}=1.58
,\end{equation}
\begin{equation}
\frac{M_{200}}{M_{500}}=1.58
.\end{equation}

\item NFW profile (Navarro et al. 1997):
\begin{equation}
\rho (r)=\frac{A}{r(r+r_s)^2}
,\end{equation}
so the mass within a radius $R_x$ is
\begin{equation}
M_x\equiv M(R_x)=4\pi \,A
\left[ln\left(1+\frac{R_x}{r_s}\right)-\frac{R_x}{R_x+r_s}\right]
,\end{equation}\begin{equation}
M_x=\frac{4}{3}\pi\,R_x^3\,x\,\rho_c(z)  
,\end{equation}
where $\rho _c(z)$ and $x$ have been defined above.

The distance $r_s$ is related to the virial radius $R_{178}$ (Lokas \& Mamon 2001) and the concentration parameter $c$ through 
\begin{equation}
r_s=\frac{R_{178}}{c}
,\end{equation}
with $c=5$ (Lokas \& Mamon 2001), and we take $H_0=70$ km s$^{-1}$ Mpc$^{-1}$, $\Omega _m=0.3$, which leads to
\begin{equation}
r_s=0.20\left(\frac{M_{178}}{10^{14}\,{\rm M_\odot }}\right)^{1/3}
\left(\frac{H(z)}{H_0}\right)^{2/3}\ \ {\rm Mpc}
,\end{equation}
and, numerically we calculate for low $z$ linear approximation
\begin{equation}
\frac{M_{200}}{M_{500}}\approx 1.40(1+z)^{0.088} 
,\end{equation}
where the mass dependence is negligible, and there is a slight redshift dependence.

\end{enumerate}

\end{document}